\documentclass[aip,jcp,amsmath,reprint]{revtex4-1}
\usepackage{txfonts}
\usepackage{dcolumn}
\usepackage{graphicx}
\usepackage{color}
\newcommand{\half}{{\frac{1}{2}}}

\begin{document}

\title{Automatic code generation enables nuclear gradient computations for fully internally contracted multireference theory}
\date{\today}
\author{Matthew K. \surname{MacLeod}}
\author{Toru \surname{Shiozaki}}
\affiliation{\mbox{Department of Chemistry, Northwestern University, 2145 Sheridan Rd., Evanston, Illinois 60208, USA}}

\begin{abstract}
Analytical nuclear gradients for fully internally contracted complete active space second-order perturbation theory (CASPT2) are reported.
This implementation has been realized by an automated code generator that can handle spin-free formulas for the CASPT2 energy and its derivatives with respect to
variations of molecular orbitals and reference coefficients.
The underlying complete active space self-consistent field and the so-called $Z$-vector equations are solved using density fitting.
The implementation has been applied to the vertical and adiabatic ionization potentials of the porphin molecule to illustrate its capability.
\end{abstract}

\maketitle

Substantial effort has been devoted to implementing complex formulas in quantum chemistry, resulting in accurate and useful computational tools for chemical applications.
However, some equations are too complicated to be handled manually.
Among such examples is the analytical nuclear gradient theory of fully internally contracted complete active space second-order perturbation theories (FIC-CASPT2 or simply CASPT2)\cite{Andersson1992JCP,Wolinski1987CPL,Pulay2011IJQC}
and its variants,\cite{Celani2000JCP,Angeli2001JCP,Angeli2002JCP}
whose complexity has hampered their implementation for more than two decades since FIC-CASPT2 was developed.\cite{Andersson1992JCP}
In this study we address this problem by employing an automatic code generation approach to help enable many chemical applications.
For instance, such implementations can be used for the geometry optimization\cite{Schlegel2011Wiley} of strongly correlated molecules
as complex as those investigated by respective single-point calculations.
They also have the potential to replace mean-field models (e.g., complete active space self-consistent field, or CASSCF)
often used in photodynamics simulations involving the ground and excited states of molecules.\cite{Levine2007ARPC} 

The challenge in implementing the nuclear gradients for the FIC-CASPT2 method can be easily recognized as follows.
The state-specific CASPT2 energy functional for the $n$-th state is
\begin{align}
&\mathcal{E} = \langle \Phi_0 | \hat{H} |\Phi_0\rangle + 2\langle \Phi_0|\hat{H}|\Phi_1\rangle +  \langle \Phi_1|\hat{f}-E^{(n)}_0|\Phi_1\rangle,
\end{align}
in which $\hat{f}$ is the standard state-specific Fock operator, and $E_0^{(n)}=\langle \Phi_0 | \hat{f} | \Phi_0\rangle$.
The reference and correlated wave functions are defined as
(using the reference CI coefficients $c^{(n)}_I$)
\begin{align}
&|\Phi_0\rangle = \sum_I c^{(n)}_I |I\rangle,\label{zero}\\
&|\Phi_1\rangle = \sum_\Omega T_\Omega \hat{E}_\Omega |\Phi_0\rangle =  \sum_\Omega \sum_I T_\Omega \hat{E}_\Omega c^{(n)}_I |I\rangle,
\label{first}
\end{align}
where $I$ labels Slater determinants, and
$\hat{E}_\Omega$ are external and semi-external excitation operators (such as $\hat{E}_{ai,bj}$ and $\hat{E}_{rs,at}$).\cite{Andersson1992JCP}
In this article $a$ and $b$ label virtual orbitals, $i$ and $j$ label closed orbitals, $r$, $s$, $t$, and $u$ label active orbitals, and $x$, $y$, $z$, and $w$ label any orbitals.
The energy functional $\mathcal{E}$ is minimized with respect to the $T_\Omega$ amplitudes to give the CASPT2 energy.
The nuclear energy gradients are the total derivative of the energy with respect to nuclear displacements $R$:
\begin{align}
\frac{d \mathcal{E}}{dR} &= \mathcal{E}\left(\frac{\partial\hat{H}}{\partial R}, T_\Omega, \mathbf{C}, c^{(l)}_I\right)\nonumber\\
      &+\sum_\Omega\frac{d T_\Omega}{d R}\frac{\partial}{\partial T_\Omega}\mathcal{E}\left(\hat{H}, T_\Omega, \mathbf{C}, c_I^{(l)}\right)  \nonumber\\
      &+\sum_{rstu} \frac{d \kappa_{rs}}{d R} \frac{\partial C_{tu}}{\partial \kappa_{rs}}\frac{\partial}{\partial {C}_{tu}}\mathcal{E}(\hat{H}, T_\Omega, \mathbf{C}, c^{(l)}_I)\nonumber\\
      &+\sum_m \sum_I \frac{d c^{(m)}_I}{d R}\frac{\partial}{\partial c^{(m)}_I}\mathcal{E}(\hat{H}, T_\Omega, \mathbf{C}, c^{(l)}_I),
\label{pert}
\end{align}
in which $\mathbf{C}$ is the molecular-orbital coefficient matrix parameterized by an anti-Hermitian matrix $\boldsymbol{\kappa}$,
\begin{align}
\mathbf{C} =\mathbf{C}_\mathrm{init} \exp(\boldsymbol{\kappa}).
\end{align}
The first term on the right hand side of Eq.~\eqref{pert} is the Hellmann--Feynman force,\cite{Helgakerbook}
and the second term is zero in the simplest state-specific CASPT2 case because $\mathcal{E}$ is stationary with
respect to variation of $T_\Omega$.
The third term also appears in the standard single-reference algorithms.
The complexity of the equations for FIC-CASPT2 nuclear gradients stems from the last term, which is associated with the reference-coefficient derivatives;
since the first-order wave functions are expanded in a basis that is dependent on $c^{(n)}_I$ [Eq.~(\ref{first})], every single term in $\mathcal{E}$ contributes to $\partial \mathcal{E}/\partial c^{(n)}_I$ in a nontrivial way.

The use of partially internally contracted or uncontracted basis functions\cite{Werner1988JCP} (referred to as WK-CASPT2 in the following) for first-order wave functions greatly simplifies the equations,
making it tractable to manually implement nuclear gradients for such variants.\cite{Nakano1998JCP,Dudley2003JCP,Celani2003JCP,Shiozaki2011JCP3,Gyorffy2013JCP}
This is because, in these methods, parts (or all) of the first-order wave function are expanded in terms of excited Slater determinants,
\begin{align}
\sum_\Omega \sum_I T_{I,\Omega} \hat{E}_\Omega |I\rangle,
\end{align}
which are not dependent on $c^{(n)}_I$ (note that in practice only distinct determinants are included in the sum).
There is also an implementation of nuclear gradients of uncontracted multireference configuration interaction (MRCI).\cite{Shepard1987IJQC}
However, the formal scaling of the size of first-order wave functions in these methods is factorial with respect to the number of active orbitals,
rendering them sub-optimal for large calculations.\cite{Celani2000JCP}

Over the last decade, various automatic code generation approaches have been developed to replace tedious, error-prone manual implementation processes in quantum chemistry.\cite{Hirata2006TCA}
The automated higher-order coupled-cluster (CC) implementations by K\'allay {\it et al.}\cite{Kallay2001JCP}
and by Hirata\cite{Hirata2003JPCA,Hirata2004JCP} were the first to demonstrate that
the automation strategy can produce programs that are competitive with hand-optimized codes in serial and massively parallel environments, respectively.  
Hanrath and co-workers have realized an arbitrary-order CC code generator that is almost optimal.\cite{Hanrath2010JCP, Engels-Putzka2011JCP}
Recently this strategy has been extended to various methods such as CC-F12,\cite{Shiozaki2008PCCP,Shiozaki2008JCP,Kohn2008JCP,Shiozaki2009JCP} relativistic CC,\cite{Nataraj2010JCP}
local CC,\cite{Rolik2011JCP,Kats2013JCP} open-shell CC,\cite{Datta2013JCTC,Datta2014JCP} and excited-state CC methods.\cite{Hirata2004JCP,Wladyslawski2005AQC}
Note that these automation techniques are complementary to meta-language approaches
(such as {\sc sial},\cite{WCMS:WCMS77} {\sc itf},\cite{Sham2011JCP} 
{\sc libtensor},\cite{libtensor} {\sc tiledarray},\cite{tiledarray}
to name a few), because code generators can be used to synthesize computer codes in any meta language as well.

Furthermore the automation strategy has been applied to development of multireference electron correlation theories.
Neuscamman {\it et al.} used an automated scheme for the canonical transformation theory;\cite{Neuscamman2009JCP}
Parkhill {\it et al.} developed local active-space methods (e.g., an active-space perfect quadruples model\cite{Parkhill2009JCP})
using a sparse framework;\cite{Parkhill2010MP}
Hanauer and K\"ohn extended their string-based code to implement an internally contracted MRCC method.\cite{Hanauer2011JCP}
Saitow {\it et al.} reported a fully internally contracted MRCI method based on density-matrix-renormalization-group reference functions.\cite{Saitow2013JCP}

In this work we extend the automatic code generation approach to realize FIC-CASPT2 nuclear gradients that have been sought for a long time.
Following the standard approach,\cite{Celani2003JCP,Helgakerbook}
we use the CASPT2 Lagrangian and the so-called $Z$-vector equation,\cite{Handy1984JCP} instead of directly evaluating Eq.~(\ref{pert}),
to avoid computation of geometry derivatives of wave-function parameters (such as $d c^{(m)}_I/d R$).
The state-specific CASPT2 Lagrangian is\cite{Celani2003JCP} 
\begin{align}
\label{lagrangian}
\mathcal{L} &= \mathcal{E}
+ \half \mathrm{tr}[\mathbf{Z}(\mathbf{A}-\mathbf{A}^\dagger)]
        -\half \mathrm{tr}[\mathbf{X}(\mathbf{C}^\dagger \mathbf{S} \mathbf{C}-\mathbf{I})] \nonumber\\
&\quad + \sum_m W_m \sum_{IJ} z^{(m)}_J \left[\langle J | \hat{H}|I\rangle -(E^{(m)}_{0}+E^{(m)}_1) \delta_{IJ}\right] c^{(m)}_I\nonumber\\
&\quad  - \half \sum_m W_m x_m\left[\sum_I (c^{(m)}_I)^2-1\right] + \sum_i^\mathrm{core}\sum_j^\mathrm{closed}z_{ij}f^{sa}_{ij}.
\end{align}
Each term on the right hand side (other than $\mathcal{E}$) corresponds to a constraint arising from the CASSCF reference calculation.
$\mathbf{A}$ is the orbital gradient of CASSCF, and $\mathbf{Z}$ is its Lagrange multiplier.
$\mathbf{S}$ is the overlap matrix, and $\mathbf{X}$ is the Lagrange multiplier for orbital orthogonality.
The next two terms are related to the stationary condition in the full configuration interaction in the active space performed in CASSCF.
Here $m$ labels states averaged in CASSCF and $W_m$ is the weight in the state averaging.
The final term originates from the use of the frozen core approximation in CASPT2.
See details in Ref.~\onlinecite{Celani2003JCP}.
When the stationary conditions on $\mathcal{L}$ with respect to all the parameters and multipliers are met, the nuclear gradients can be computed as
\begin{align}
\frac{d \mathcal{E}}{dR} = \frac{\partial \mathcal{L}}{\partial R}
= \mathcal{L}\left(\frac{\partial \hat{H}}{\partial R}, \frac{\partial \hat{S}}{\partial R}, \cdots\right),
\end{align}
since only molecular integrals have explicit dependence on the nuclear displacement $R$.
We will consider level shifts\cite{Roos1995CPL} in future work, which requires the additional implementation of the so-called $\lambda$ equation (as do multistate variants\cite{Shiozaki2011JCP3}).

We have developed a new automated code generator, named {\sc smith3}, that derives equations on the basis of the spin-free version of Wick's theorem,
in which the normal ordering is defined with respect to the closed-Fock vacuum.
In addition, {\sc smith3} expresses the terms that involve active-orbital indices as a sum of canonical terms
so that they can be computed from canonical density matrices and its derivatives, e.g.,
\begin{subequations}
\begin{align}
&\langle \Phi_0 | \sum_{\rho \sigma}r_\sigma s^\dagger_\sigma t^\dagger_\rho u_\rho|\Phi_0\rangle = 2\delta_{rs}(\Gamma_0)_{tu} - \delta_{rt}(\Gamma_0)_{su}- (\Gamma_0)_{sr, tu},\\
&\langle I | \sum_{\rho \sigma} r_\sigma s^\dagger_\sigma t^\dagger_\rho u_\rho|\Phi_0\rangle = 2\delta_{rs}(\Gamma_0)^I_{tu} - \delta_{rt}(\Gamma_0)^I_{su}- (\Gamma_0)^I_{sr, tu},
\end{align}
\end{subequations}
with $\sigma$ and $\rho$ labeling spins.
Note that $(\Gamma_0)_\Lambda = \langle \Phi_0|\hat{E}_\Lambda|\Phi_0\rangle$ and $(\Gamma_0)_\Lambda^I = \langle I|\hat{E}_\Lambda|\Phi_0\rangle$ where $\hat{E}_\Lambda$ is a general operator.
Here {\sc smith3} is used to implement the following expressions:
\begin{subequations}
\begin{align}
&\langle \Psi| \hat{E}^\dagger_\Omega \hat{\mathcal{G}} \hat{\mathcal{R}}|\Psi \rangle, \label{gen1}\\
&\langle \Psi | \hat{\mathcal{R}}^{\prime\dagger} \hat{E}_\Lambda \hat{\mathcal{R}} | \Psi\rangle, \label{gen2}\\
&\langle I | \hat{\mathcal{R}}^{\prime\dagger} \hat{\mathcal{G}} \hat{\mathcal{R}} | \Psi\rangle, \label{gen3}
\end{align}
\end{subequations}
in which $|\Psi \rangle = \sum_I t_I |I\rangle$ is any multi-configuration reference function,
and $\hat{\mathcal{G}} = \sum_\Lambda G_\Lambda \hat{E}_\Lambda$ and $\hat{\mathcal{R}} = \sum_\Omega R_\Omega \hat{E}_\Omega$ are general and excitation operators, respectively.
Note that the determinant index $I$ is treated analogously to the orbital indices in the generated code; for instance, $(\Gamma_0)^I_{tu}$ is viewed as a three-index tensor whose size is $n_\mathrm{det}n_\mathrm{act}^2$ ($n_\mathrm{det}$ and $n_\mathrm{act}$ are the numbers of the determinants and the active orbitals, respectively).

Using this machinery, we automate the nuclear-gradient implementation  as follows. 
First, to optimize $T_\Omega$ the program for computing the CASPT2 residual vectors is generated using Eq.~\eqref{gen1}, i.e.,
\begin{align}
\frac{\partial \mathcal{L}}{\partial T_\Omega}  =2 \left[\langle \Omega | \hat{f}-E^{(n)}_0|\Phi_1\rangle + \langle \Omega | \hat{H} | \Phi_0\rangle\right].
\end{align}
Second, the $Z$-CASSCF equation is solved, which is a set of coupled equations defined as
\begin{subequations}
\begin{align}
&\frac{\partial \mathcal{L}}{\partial c^{(n)}_I} = 0, \label{zci}\\
&\frac{\partial \mathcal{L}}{\partial \kappa_{rs}} = 0. \label{zmo}
\end{align}
\end{subequations}
For Eq.~(\ref{zci}) the reference-coefficient derivatives of the CASPT2 energy,
\begin{align}
y^{(n)}_I = \frac{\partial \mathcal{E}}{\partial c^{(n)}_I},\label{smally}
\end{align}
are implemented by {\sc smith3} using Eq.~(\ref{gen3}).
Next, for Eq.~\eqref{zmo} the MO-coefficient derivatives of the CASPT2 energy, 
\begin{align}
Y_{rs} = \frac{\partial \mathcal{E}}{\partial \kappa_{rs}},
\end{align}
are calculated from the one- and two-body density matrices (see Refs.~\onlinecite{Celani2003JCP,Gyorffy2013JCP} for explicit formulas).
The density matrices are defined as
\begin{subequations}
\begin{align}
&(\Gamma_1)_{xy} = 2\langle \Phi_0| \hat{E}_{xy} | \Phi_1\rangle,\\
&(\Gamma_2)_{xy} = \langle \Phi_1| \hat{E}_{xy} |\Phi_1\rangle - \langle \Phi_0|\hat{E}_{xy}|\Phi_0\rangle\langle\Phi_1|\Phi_1\rangle,\\
&(\Gamma_1)_{xy,zw} = 2\langle \Phi_0| \hat{E}_{xy,zw} | \Phi_1\rangle.
\end{align}
\end{subequations}
and are implemented using Eq.~(\ref{gen2}).
Given $y^{(n)}_I$ and $Y_{rs}$, solutions of $Z$-CASSCF [$\mathbf{Z}$, $\mathbf{X}$, $z^{(m)}_I$, and $z_{ij}$ in Eq.~\eqref{lagrangian}] can be obtained
as detailed in Ref.~\onlinecite{Celani2003JCP}. 
Using these wave function parameters and the Lagrange multipliers, effective density matrices are formed, which are then
contracted to two-index and three-index gradient integrals.\cite{Celani2003JCP,Gyorffy2013JCP}

The generated code uses a tile-based data layout that is similar to those used in {\sc tce}\cite{Hirata2003JPCA}
and in the earlier version of {\sc smith}.\cite{Shiozaki2008PCCP,Shiozaki2008JCP}
All the code implemented in the {\sc bagel} package\cite{bagel} and the code generator {\sc smith3}\cite{smith} are openly available under the GNU General Public License.
The technical details on the implementation, working equations, and source code of the {\sc smith3} program are also found in Supplementary Materials.\cite{supp}
CASSCF and $Z$-CASSCF were manually implemented in {\sc bagel} using density fitting (DF) for efficiency as reported in Ref.~\onlinecite{Gyorffy2013JCP}.
In CASPT2, four-index two-external integrals were explicitly constructed from DF integrals. 
The {\sc smith3} program generated ca.~1150 tasks, the majority of which are tensor contractions.
Each task is expressed as a node of a tree-like directed acyclic graph, which we traverse at runtime.
This infrastructure should assist in interfacing {\sc smith3} code to parallel-runtime libraries in the future. 

First, to show the numerical impact of full internal contraction on geometrical parameters,
we optimized the ground-state geometry of the $N$,$N$'-diiminato-copper-dioxygen complex [(H$_5$C$_3$N$_2$)CuO$_2$] in its side-on coordination configuration.
The ground state is singlet.
We used the (14$e$, 9$o$) active space consisting of an in-plane $d$ orbital of copper and eight valence orbitals of dioxygen
as suggested in earlier work.\cite{Sham2011JCP,Gyorffy2013JCP}
The aug-cc-pVDZ\cite{Dunning1989JCP,Kendall1992JCP} and def2-QZVPP/JKFIT\cite{Weigend2008JCompC} basis sets were used for orbital and auxiliary functions, respectively.
Table~\ref{cuo2} compiles optimized Cu--O and O--O bond lengths.
It is evident that neither the degrees of internal contraction (i.e., CASPT2 and WK-CASPT2)
nor the DF approximation has impact on the bond lengths. 
Our program did not take advantage of spatial symmetry.
One optimization step took about 13~min. using 2 Xeon E5-2650 CPUs (2.0~GHz). More than half the time is spent for evaluation of Eq.~\eqref{smally}.
Note, however, that the code generated by {\sc smith3} has not been threaded efficiently,
and there is room for further improvement.

\begin{table}
\caption{Optimized Cu--O and O--O bond lengths (in {\AA}) for the ground state of (H$_5$C$_3$N$_2$)CuO$_2$ 
using CASSCF and CASPT2 with aug-cc-pVDZ and the (14$e$, 9$o$) active space. DF was used unless otherwise stated.\label{cuo2}}
\begin{ruledtabular}
\begin{tabular}{lcc}
Method  & Cu--O & O--O  \\\hline
CASSCF  & 1.886 & 1.386 \\
CASPT2  & 1.820 & 1.399 \\
WK-CASPT2\footnotemark[1] & 1.820 & 1.400 \\ 
WK-CASPT2 (w/o DF)\footnotemark[1] &  1.820 & 1.400 \\ 
\end{tabular}
\end{ruledtabular}
\footnotetext{Partially contracted CASPT2 computed using {\sc molpro}.\cite{molpro}}
\end{table}

Next we calculated the vertical and adiabatic ionization potentials (IPs) of the porphin molecule (C$_{20}$H$_{14}$N$_4$) using the optimized geometries computed by CASPT2. 
The cc-pVDZ basis set\cite{Dunning1989JCP} was used together with the corresponding JKFIT basis set\cite{Weigend2002PCCP} for DF.
The (4$e$, 4$o$) and (3$e$, 4$o$) active spaces were used,\cite{supp} which consist of the four frontier orbitals of Gouterman's model.\cite{Gouterman1959JCP}
The numbers of (correlated) inactive and virtual orbitals were 55 and 323, respectively.
One optimization step took about 30 min. on the same hardware.
For comparison, we also computed the IPs from the PBE functional\cite{Perdew1992PRL}
and MP2 (with an unrestricted variant for the radical cation) using {\sc turbomole}.\cite{Furche2014WIREs}
The PBE calculations were performed using the def2-SVP basis set.\cite{Weigend2005PCCP}
Geometry optimization using all methods including CASPT2 was performed without imposing spatial symmetry;
the geometry of the neutral porphin was found to belong to the $D_{2h}$ symmetry group,
whereas that of the radical cation was found to be $C_{2h}$
(even when an initial geometry was set to a $D_{2h}$ structure, optimization converged to this minimum). 
Similar symmetry breaking of metalloporphyrin cation radicals due to the pseudo Jahn--Teller effect
has been reported in the literature.\cite{Vangberg2002JACS}
The optimized geometry of the radical cation from unrestricted PBE was found to be $D_{2h}$.
We were not able to optimize the geometry of the radical cation using unrestricted MP2 due to wave function instability.
The geometrical parameters are compiled in Supplementary Materials.\cite{supp}
The IPs are shown in Table~\ref{iptable}. 
The vertical IP computed by CASPT2 (6.84~eV) is in good agreement with the experimental value (6.9~eV).\cite{Dupuis1980CPL}
The difference between the vertical and adiabatic IPs computed at 0.18~eV
is an order of magnitude larger than that computed using PBE.
Furthermore we computed the IPs with CASPT2 using the PBE-optimized geometries. 
As expected, the vertical IP is almost identical to that computed at the CASPT2 geometry;
however, the adiabatic IP is larger than the vertical IP, which 
attests to the importance of geometry optimization at the CASPT2 level.

\begin{table}[t]
\small
\caption{Ionization potentials (eV) of the porphin molecule. The cc-pVDZ basis set was used unless otherwise stated. The (4$e$, 4$o$) and (3$e$, 4$o$) active spaces were used in the CASPT2 calculations.\label{iptable}}
\begin{ruledtabular}
\begin{tabular}{ldddd}
Method & \multicolumn{1}{c}{Vertial IP}  &  \multicolumn{1}{c}{Adiabatic IP} &  \multicolumn{1}{c}{$\Delta$IP} & \multicolumn{1}{c}{$\langle S^2\rangle$\footnotemark[1]}\\
\hline
PBE\footnotemark[2]  &   6.70     &  6.68 & 0.02 &  0.77 \\
MP2                  &   8.51     &  ...\footnotemark[3]   &  ...     &  1.47 \\ 
CASPT2               &   6.84     &  6.65  & 0.18 &  0.75 \\
CASPT2/PBE           &   6.83     &  6.85  & -0.02 & 0.75 \\ 
Experiment\footnotemark[4] & 6.9 & ... & ... & ... 
\end{tabular}
\end{ruledtabular}
\footnotetext[1]{$\langle S^2\rangle$ of the radical cation at the equilibrium geometry of the neutral.}
\footnotetext[2]{Computed using the def2-SVP basis set.}
\footnotetext[3]{Due to instability we could not optimize the geometry of the radical cation.}
\footnotetext[4]{Taken from Ref.~\onlinecite{Dupuis1980CPL}}
\end{table}

In summary we have used automatic code generation to realize analytical CASPT2 nuclear gradients with full internal contraction. 
Our implementation has been applied to the $N$,$N$'-diiminato-copper-dioxygen complex to show that errors due to full internal contraction are marginal. 
We have also computed the vertical and adiabatic IPs of the porphin molecule to illustrate the capability of our implementation.
There is, however, room for improvement in our program in terms of efficiency and storage requirement;
currently, application of our program is limited by the storage of
\begin{align}
(\Gamma_0)^I_{rr',ss',tt'},\quad     (\tilde{\Gamma}_0)^I_{rr',ss',tt'}=\sum_{uu'} (\Gamma_0)^I_{rr',ss',tt',uu'} f_{uu'},
\end{align}
and the $T_\Omega$ amplitudes of $n_\mathrm{occ}^2n_\mathrm{bas}^2$ size ($n_\mathrm{occ}$ and $n_\mathrm{bas}$ are the numbers of occupied orbitals and basis functions, respectively).
Further optimization and distributed-memory parallelization of our program are warranted and will be performed in the future.
We will also consider the level shift,\cite{Roos1995CPL} other zeroth-order Hamiltonians,\cite{Ghigo2004CPL,Angeli2002JCP} and multistate extensions.\cite{Finley1998CPL,Shiozaki2011JCP3,Granovsky2011JCP}

The debugging of parts of the code in {\sc bagel}, including $Z$-CASSCF, was facilitated by existing implementations in {\sc molpro}.\cite{molpro}
This work has been supported by Department of Energy, Basic Energy Sciences (Grant No.~DE-FG02-13ER16398) and the Air Force Office of Scientific Research Young Investigator Program (Grant No.~FA9550-15-1-0031).

\bibliography{spinfree}
\vfill

\end{document}